# Documentless Assessments Using Nominal Group Interviews

Eduardo Miranda

Carnegie Mellon University, Institute for Software Research
Pittsburgh, Pennsylvania, USA
mirandae @ andrew . cmu . edu

**Abstract.** This paper describes a group interview technique designed to support document-less process assessments while promoting at the same time collaboration among assessment participants. The method was successfully used in one consulting assignment were it got previously discording participants, talking to each other and agreeing on the issues. The technique borrows from agile software development the concept of user stories to cast CMMI's specific practices in concrete terms and the Planning Poker technique, instead of document reviews and audit like interviews, for fact finding and corroboration.

## 1      Introduction

The group interview technique presented in this paper was developed by the author to support the assessment portion of a process improvement initiative launched by the management of a research agency which, as part of its mandate, develops and maintains a very sophisticated application used by more than 2,000 scientists all over the world. The organization was aware of its two main problems concerning this application: the accumulation of technical debt resulting from the development of features over a period of ten years without much architectural oversight and little refactoring, and the lack of a common development process fueled by the internal dissent of highly specialized and almost irreplaceable specialists. A previous attempt to address these problems had backfired due to the peremptory approach followed by the person responsible for the improvement initiative. In requesting an assessment of their current ways of working, management had two objectives in mind: pinpointing specific problems by means of a recognized best practices framework and getting the development group to buy-in into the initiative. The development group, which consisted of about 25 software engineers and 6 subject matter experts, was skeptical of what was perceived as a bureaucratic exercise getting in the way of doing the work.

In this context, a Standard CMMI Appraisal Method for Process Improvement (SCAMPI) like assessment based on document reviews and audit-like interviews was out of the question. In the opinion of the author, this approach would not only had met with the passive resistance of those involved but would have further convinced them, that they were right in their rejection of the whole process.

Through his teaching activities in the Master of Software Engineering at Carnegie Mellon University, the author had learned first-hand about the power of user stories to synthetize a lot of information in a concise format and that of the Planning Poker to get people talking and helping them to arrive to a consensus. So he thought to himself: why not use them for fact finding and corroboration? Both techniques looked apt for the job and would give the assessment a much needed fresh look in the eyes of the developers.

The assessment comprised individual interviews with managers and user representatives and two group interviews with practitioners at different locations. The interviews with managers and user representatives had for purpose finding out the pain points, the improvement goals, the degree of support for the initiative and any impediments they saw moving forward. The group interviews with practitioners focused on the state of the practice within the group vis-à-vis all level 2 and some level 3 process areas of the CMMI, the issues from the practitioners' point of view and whether the group had a congruent view of the problems and their possible solution. **Fig. 1** depicts the group interview process, the focus of this experience report.

The proposed group interview technique could be easily used in the context of other lightweight assessment processes such as the ADEPT [1] and the Modular Mini-Assessment [2] methods.

The rest of the paper is organized as follows: Section 2 provides cursory information on the techniques employed: user stories, the Nominal Group Technique, SCAMPI and the Planning Poker. Section 3 explains why and how to express specific practices as user stories. Section 4 discusses the modified Planning Poker technique used in the assessment of the current practice. Section 5, describe the preparation of the final findings documents, Section 6 briefly describes management interviews and Section 7 the experience applying the method.

## 2      User stories, the Nominal Group Technique, SCAMPI and the Planning Poker

This section provides an introduction to the four techniques in which the proposed assessment method is based. Readers familiar with them may skip it.

**User stories**

In agile software development a user story [3] is an expression of a need or want of a stakeholder as well as a unit of planning.

In it most complete form, a user story includes: its title, a description, optional details called a "conversation" and a verification criteria called the confirmation [4]. Not everybody uses the four elements and not everybody calls them the same.

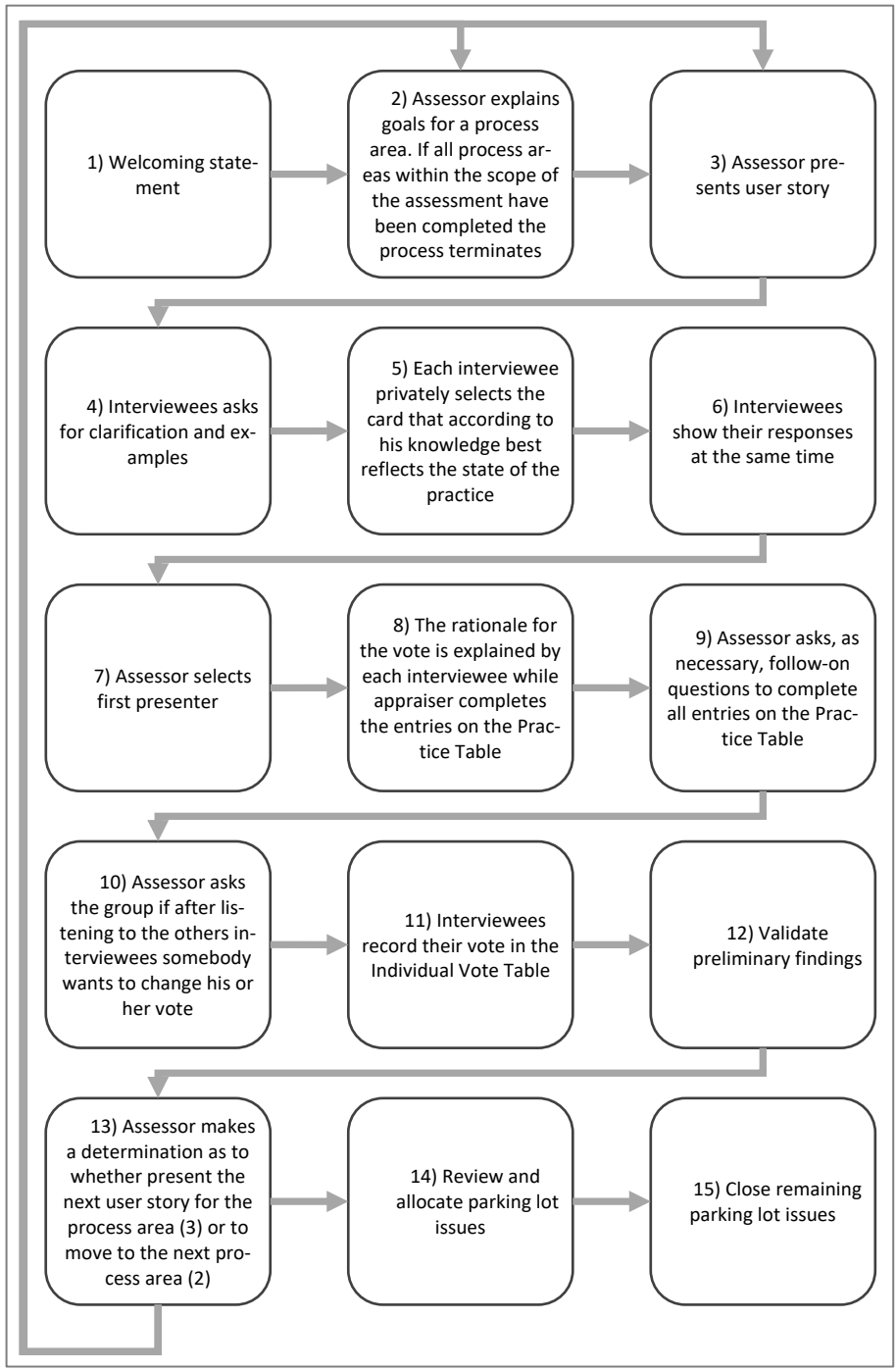

**Fig. 1.** Group interview process

The title of the user story is just that, a label by which it is known. It is commonly written as a verb phrase but some people use numbers or codes as identifiers.

The description of the user story corresponds to a need or want the system of interest has to satisfy. It is typically written using the formula: *As a <role> I want to <action> so that <benefit>*. The Role – represents who is performing the action or who benefits from it. The Action – represents the activity, function, work, or service to be provided by the system. The benefit – is the reason for performing the action. The user stories' description is written, changes and additions apart, during the project formulation.

The conversation part of the user story elaborates on the story's description. Conversations might be documented or not, but obviously any detail known or decision made at the beginning of the project that is not be written down, at least as a note attached to the story description, is condemned to be forgotten halfway through it. The conversation part of a user story could be elaborated at any time, but needs to be concluded before the team starts developing it.

The confirmation part of the story, also called the acceptance criteria, needs to be written before the team tackles its development and represents the conditions of satisfaction that will be applied to determine whether or not the story as implemented fulfills the intent as well as the more detailed requirements expressed in the conversation part.

In the context of our proposal we will only use the description part of the user story. How and why will be fully explained in Section 3.

**Nominal Group Technique**

The Nominal Group Technique (NGT) is a systematic approach for soliciting and pooling individual inputs into a group decision or assessment. The technique was developed by Delbeq and Van de Ven [5] in the late sixties. NGT combines both individual and group phases. A typical NGT meeting proceeds as follows:

1. Private generation of ideas
2. Public display of all ideas
3. Serial clarification/explanation of each idea
4. Preliminary vote on all the ideas
5. Serial explanation of each individual's preliminary vote
6. Private reflection on all the explanations listened to
7. Final vote
8. Aggregation of results

The introduction of individual and group phases combined with the serial presentations mechanism allows for a variety of viewpoints, knowledge, and interests to inform the decision while preventing more confident, outspoken, or higher status members from dominating the discussion.

**Planning Poker**

The Planning Poker, proposed by J. Grenning [6] and popularized by M. Cohn [7] is an estimation technique that as the NGT relies on alternative individual and group phases to avoid bias and conformity effects while maximizing the number of inputs that inform the decision. The steps in the planning poker are:

1. The product manager explains a user story to the team
2. The development team ask for clarification
3. Private preliminary estimation
4. Public display of all preliminary estimates
5. Serial explanation of each preliminary estimate
6. Private reflection on all the explanations listened
7. Final individual estimate
8. Aggregation of results

One of the interesting things about the planning poker, is that estimates are made along a predefined scale printed on cards and hidden from other participants until showdown, see **Fig. 2**, and hence its name. Having witnessed very dull electronic mediated planning poker sessions, the author believes it is the cards and the showdown that appeal to our ludic mind and making it fun to participate and heightening participants' engagement.

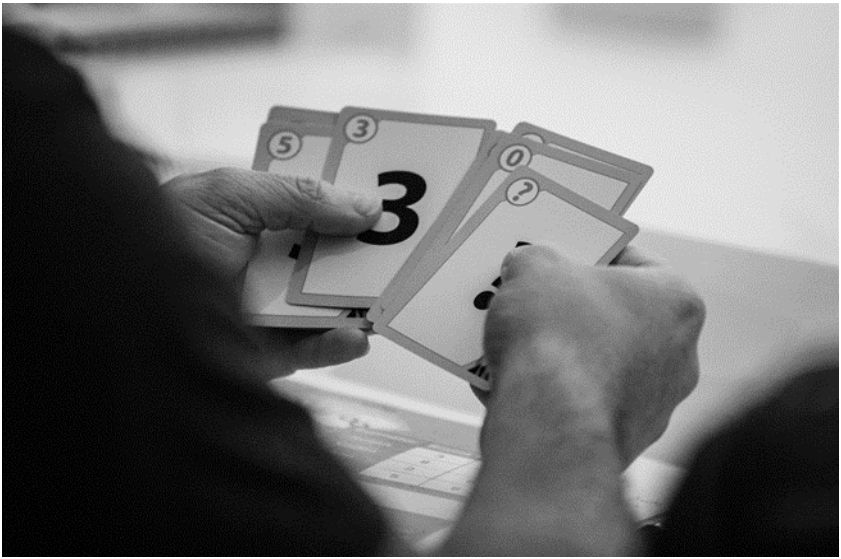

**Fig. 2** Software developer privately estimating a user story, Mountain Goat software, 2014

For the purpose of process assessment the author designed the set of cards shown in **Fig. 3**. Its use will be explained in Section 4.

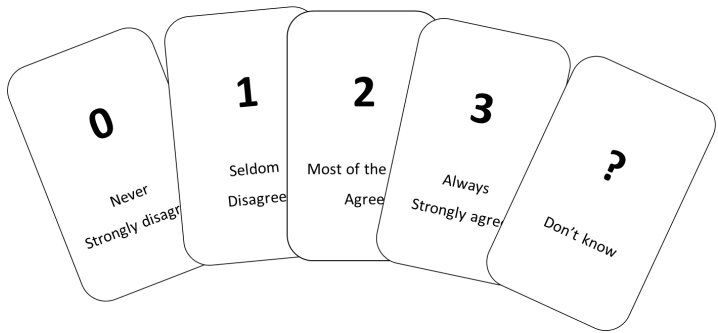

**Fig. 3.** Voting cards. Notice the cards do not include a neutral option. This was purposefully designed to force participants to take a stand on one side or other of the scale. A similar scale was proposed by [2]

**SCAMPI**

The <u>S</u>tandard <u>C</u>MMI <u>A</u>ppraisal <u>M</u>ethod for <u>P</u>rocess <u>I</u>mprovement, is in reality a family of appraisal methods developed by the Software Engineering Institute [8]. There are basically three variations of SCAMPI known as "A", "B" and "C". The purpose of SCAMPI A is to establish an official rating with regards to the assessed organization's maturity while the "B" and "C" versions focus on process improvement. The variants differ mostly in the level of corroboration required to verify whether an organization performs or not a certain process and in the qualifications required of the appraiser. For example, a SCAMPI "A" has to be led by a certified assessor while SCAMPIs "B" and "C" can be performed by a trained and experienced individual.

SCAMPI appraisals are based on the verification and validation of the objective evidence: direct work products, indirect work products and affirmations, provided by the appraised organization to identify strengths and weaknesses relative to the CMMI.

**Table 1** below compares the requirements for data collection, consolidation, and validation [9] for the different classes of assessment against those of the proposed method.

**Table 1.** Comparison of SEI's appraisal requirements with the attributes of the proposed method

| Requirement | A | B | C | Proposed method |
|---|---|---|---|---|
| Collect affirmation data (e.g. by conducting interviews with project or work group leaders, managers, and practitioners). | x | x | | Yes. Through the group and management interviews |
| Collect data by reviewing artifacts (e.g., organizational policies, project or work group procedures, slides from past presentations, and implementation-level work products). | x | x | | No. Corroboration is achieved through a voting procedure. |
| Validity of findings and ratings are done by consensus. | x | x | x | Not applicable. The assessment is performed by a single individual. |
| Consolidation of collected data during an appraisal meets the following criteria:<br>a. Findings are derived from evidence seen or heard during data collection sessions.<br>b. Findings are clearly worded, phrased without attribution, and expressed in terminology used by the staff working in the organizational unit.<br>c. Evidence supporting the finding is traceable to the project, work group or organizational unit.<br>d. Findings are relevant to the appraisal reference model and can be associated with a specific model component. | x | x | x | Mostly. Individual votes are consolidated according to the vote interpretation rules. There is not traceability (c) to organizational unit or project. |
| Findings are verified according to the following criteria:<br>a. The finding is based on corroborated evidence.<br>b. The finding is consistent with other verified findings. | x | x | | Yes. Through the voting mechanism and the explanations provided by the assessment participants. |
| Corroboration of evidence must satisfy the following criteria:<br>a. The evidence is obtained from at least two different sources.<br>b. At least one of the two sources must reflect the work that is actually being done (e.g., process area implementation). | x | x | | Affirmations are corroborated or not by the voting procedure. |
| Collect sufficient data to cover the scope of the appraisal. | x | x | x | Yes. Completion of the Practice Table. |
| Findings are validated with appraisal participants | | x | | Yes. Findings are validated during steps 12, Validate Preliminary Findings and 15, Close Remaining Parking Lot Issues, of the group interview process. |

## 3 Expressing specific practices as user stories.

Assessing the group's way of working against the process areas of the CMMI requires verifying whether the practices defined by it are performed or not, and in the affirmative if they do so in an effective and efficient manner. To do this, the group interview process presented in the next section walks assessment participants through all the practices in scope, asking them whether the practice is implemented or not, and whether they find it valuable. The participants' answers and more importantly, their buy-in into the process depends a lot on how the question is formulated [10]. For example, while few people will argue that connecting test cases to the functionality they verify is an important quality of a software development process, asking if they *"maintain bidirectional traceability among requirements and work products"* would rise quite a number of eyebrows.

Of course the two phrases are not equivalent, the first is an instance of the second and is limited to a single work product. The point here is that while CMMI rightly aims for generality, response accuracy, buy-in and the development of a shared understanding is built around specific and not abstract constructs. The situation has been clearly described by Arent et al [11] in the recount of their experience at Ericsson: *"The problem was that the project managers didn't understand the reasons for using CMM until they had actually tried to use it, and they didn't use it because they didn't understand the reasons for it. It was a vicious circle, making it difficult to succeed"*.

In our approach, to make CMMI specific practices concrete, we chose to use a slightly modified user story format: "As a <role> <personal pronoun> <practice instance> so <benefit>". This was a good vehicle for moving from the abstract to the concrete not only because most developers were already familiar and well predisposed to them but because they include who does the work or who benefits from the practice: the <role>; what is done: the <practice instance> and the reason for doing it: the <benefit>. The <personal pronoun> is just that, its only function is to make the user story grammatically correct.

The ideal user stories would be site specific. They would be crafted by the assessor using his or her CMMI knowledge as well as borrowing vocabulary and practices from the organization under assessment. **Table 2** below provides some examples as to how these user stories could look like.

Notice that there could be more than one <role> or <benefit> associated with a single <practice instance>, for example a <practice instance> could benefit or be performed by developers and testers and/or there could be multiple <benefit>s accruing from it, but in order to keep things simple we circumscribe the user story to direct performers and beneficiaries or, if already in use by the organization, a more encompassing category such as "team member", but we do not create artificial roles for the sake of economy of expression. Similarly, we limit the description of the user story to one or two direct benefits since these are all it is needed to justify a practice. Conversely, if we could not find any beneficiary for doing something, we should consider dropping the practice from the assessment, otherwise seems like the organization has to do things for the sake of the model and not for the quality of the product or to better their way of working.

**Table 2.** Recasting CMMI's specific practices as user stories. Selected examples.

| Reference | CMMI Practice | User story |
|---|---|---|
| REQM 1.3 L2 | Manage changes to requirements as they evolve during the project | As a team member I can find how user stories have evolved over time as well as their current status so I can better understand stakeholders' needs and avert "he said, she said" situations |
| PP 1.2 L2 | Establish estimates of work product and task attributes | As a team we establish estimates for user stories and tasks so that we can make commitments to our stakeholders and plan our work |
| PMC 1.1 L2 | Monitor actual values of project planning parameters against the project plan | As a team we track rate of work completion using iteration and release burn down charts so that we can keep all stakeholders abreast of our progress |
| MA 2.3 L2 | Manage and store measurement data, measurement specifications, and analysis results | As an organization we preserve our defect and velocity data so it can be used by other projects to check their initial estimates against what has been achieved and to find organizational quality issues and bottlenecks |
| RSKM 1.1 L3 | Determine risk sources and categories | As a team we have at our disposal a list of risks sources that can help us identify what might go wrong in a project and decide what to do about it |
| RSKM 2.1 L3 | Identify and document risks | As a team we make a conscious effort to identify and document potential problems so we don't overlook them |
| TS 1.1 L3 | Develop alternative solutions and selection criteria | As a team we discuss the characteristics a good software solution should possess and evaluate different solutions against them to avoid following a dead end path |
| VER 2.2 L3 | Conduct peer reviews | As developers we review each other code with the purpose of identifying bugs and non-compliances with our coding guidelines |
| VAL 1.2 L3 | Establish and maintain the environment needed to support validation | As a team we use a canary release strategy to get fast feedback from actual users |

The more abstract a concept is, the higher the level of interpretation required and in consequence the higher the variability in the understanding of the same [12]. This makes the choice of <practice instance> to be used in lieu of the corresponding CMMI abstract practice, a critical issue in eliciting definite answers from assessment participants.

Continuing with the idea of making things obvious, a simpler practice is preferred to a more complex one. In general, if the organization is not doing those things that give more bang for the buck it is unlikely they will do those that are at the fringes. Erring in the side of simplicity when the organization is doing something more elaborate, is not a problem because one or more participants in the interview are likely to recognize the intent of the practice and answer correctly while at the same time volunteering good information.

The previous discussion deals with specific practices, but what about, CMMI's generic goals and practices? A CMMI generic goal is one that applies to multiple process areas in the model. These goals and their associated practices deal with the institutionalization of the specific processes that is, whether the organization follows them routinely as part of doing business or not.

In the proposed method, the institutionalization of the process is assessed via the consistency of the interview responses and by the comments made by the interviewees. This will be explained in detail in the next section.

## 4 The group interview technique

**Fig. 1** above, describes the workflow used in the practitioners interview. The process is based on the nominal group technique proposed by Delbeq et al [5] and on the Planning Poker [6, 7], from which we borrow the idea of using cards to answer the interview questions, see Figure 2.

The two key activities in the nominal group technique are the private voting and the round-robin explanation mechanism. Both activities synergistically promote frankness, participation and engagement. Because private voting precludes people from knowing how the others will vote, people cannot piggyback on somebody else's explanations forever while maintaining some kind of intellectual consistency over the course of the assessment, so most participants would choose to be candid in their votes and explanations. The stipulation that all voting cards must be turned at the same time reduces conformity effects. The round robin mechanism promotes engagement by either giving everybody the opportunity, or by forcing, to expound its vote and in turn, listen to the explanations provided by others. In the words of Delbecq et al [5], the inventors of the method, *"The rather mechanical format of going to each member in turn to elicit ideas establishes an important behavior pattern. By the second or third round of idea giving, each member is an achieved participant in the group"*. We observed a similar pattern that is discussed in Section 6.

The selection of participants for the assessment is key to the credibility of findings and recommendations. The selection must ensure discipline coverage; balancing experienced personnel, who understand the organization well, with newcomers, who face the challenge of getting on board. Having a wide spectrum of participants also ensures domain coverage. To promote openness management shall be excluded from participation in these interviews. Since the method relies on the agreement or disagreement of the interviewees it is very important to have at least two representatives from the main development areas. The number of participants should be kept under ten in the interest of time. The following paragraphs detail each of the workflow steps.

1. Welcoming statement

Purpose: Put participants at ease. Explain the assessment process, its purpose, and rules.

Description: The assessor welcomes the participants informing them the reason for their selection and highlights the need for everybody's contribution despite differences in roles and seniority. The overall process is explained using a diagram similar to the one show in **Fig. 1**. Participants are provided with the deck of cards, see **Fig. 3**, they will use to take votes and made aware of the basic appraisal rules: no attribution of votes and comments, no right or wrong answers, that interviewees should answer to the best of their knowledge, and that questions might be skipped if it becomes obvious from previous responses that no new insights will be gained by asking them. The assessor will inform participants about breaks and other logistics. On a more mundane note, the author has found that cookies, coffee and a brief words by a senior manager concerning the importance of the initiative, will go a long way towards a successful meeting.

2. Assessor explains the goals of the process area

Purpose: Make participants cognizant of the intent of the process area the group is about to aboard.

Description: During this activity the assessor explains the overall intent of the process area and that he will be using scenarios, in the form of user stories, to exemplify specific practices but that the organization might be achieving the same through some other mechanism and for that reason it is very important to keep an open mind.

3. Assessor presents a user story

Purpose: Initiate the assessment of a specific practice. Make participants cognizant of the intent of the specific practice.

Description: The assessor presents a user story to the group and after explaining it asks if clarification is required. User stories for each process area are presented one at a time. The assessor will first put a slide with the user story text that will remain up until the next one is presented and read it aloud. During the presentation the assessor might remind the group that the <practice instance>, as well as the <role> and the <benefit>s presented are exemplars and that there might exist other <role>s performing it or other <benefit>s derived from it. The assessor ends the presentation by asking if the user story is understood or if further clarification is required.

4. Interviewees ask for clarification

Purpose: Provide an opportunity for interviewees to confirm their understanding of the intent of the specific practice and suitable alternatives.

Description: During this step the assessment participants ask questions with regards to the practice. Typical questions include the practice implementation, its goals and the protagonists. In responding, the assessor might resort to the original text of the specific CMMI practice to widen the perspective of the group in considering it. Time-boxing this period helps keep the conversation on point and minimizes wasted time. A good technique to prevent the conversation from drifting while remaining respectful of the speaking participant, is to acknowledge the argument and explain the point will be addressed on a coming process area or ask the group if the issue can be put in a parking lot to deal with it later.

5 & 6. Interviewees vote

Purpose: To obtain a collective view of the state of the practice.

Description: After answering all interviewee's questions, the assessor will direct the assessment participants to take a preliminary vote on whether the practice is always followed, often followed, seldom followed, never followed or to indicate they don't know using cards like the ones show in **Fig. 3**. Interviewees privately select the card that according to their knowledge best reflects the state of the practice. Once the assessor notices everybody has selected its card, it is important to allow adequate time for reflection, he will ask interviewees to show their votes at the same time. This last point is crucial to avoid distortions in individual judgments resulting from intentional or unintentional status, personality, and conformity pressures that might distort individual judgments.

7. Select first presenter

Purpose: Select the first participant to start the explanations round.

Description: This might seem as a trivial step, but in order to avoid primacy and recency effects, and for preventing a more extrovert personality to unduly dominate the meeting with his or her explanations or to liberate shy individuals from the stress of always being the first, it is important to choose a different starting participant for each round of explanations. Sometimes it might be the person that was second to the first in the previous round, other times it could be somebody with a dissenting or extreme vote, because as much as the private vote mitigates conformity effects, hearing a couple of their colleagues say the opposite of what one had in mind, might weaken the sound of a lonely voice. To avoid having participants misinterpreting this move as an affront or disregard for their opinions it is very important to explain this during the welcoming statement.

8. Explanations

Purpose: For the interviewees to share their views and knowledge. For the group to start building a shared understanding of the situation. For the assessor to collect diagnosis information.

Description: During this step the interviewees take turn to explain the rationale for their vote. No interruptions or references to other people's responses are allowed during each exposition. It is key that participants feel free to express varying points of view or to disagree. At this time the assessor has three responsibilities: pace the group in order to give time to everybody to talk, avert side conversations and argumentation among participants and to take notes. Notice that during this step the assessor does not attempt to clarify or seek additional information. Doing so might bias the explanations in a certain direction, when the goal of this process step is to cast a wide net. If at some point the explanations start to repeat and the remaining votes coincide, the assessor might ask the participants if somebody has anything new to add and otherwise go to the next step in order to save time.

For capturing the information in a structured manner and ensure completeness the author used the list shown in **Fig. 4** and referred to as "Practice Table" in the workflow.

---

- Is the practice being performed? Requires majority of respondents to agree or strongly agree
- Brief description if alternate practice
- Is it relevant? Adds value? If it were not executed something would not be accomplished, would cost more, etc.
- Efficient? The achievement of the goal requires an effort commensurate with the value of the outcome. The practice does not overlap or interfere with other practices
- Institutionalized? Does the staff receive training to perform it? Are adequate resources provided for performing it? Whenever a project is late, does the organization shortcut the practice with the excuse of saving time?
- Documented? Is there a document that mandates or describes the practice?
- Are there any noticeable strengths or weaknesses?
- Assuming that it makes sense, what prevents the practice from being implemented?
- Can anybody remember a problem in a project that can be traced back to deficiencies/lack of practice being performed?
- Additional comments

---

**Fig. 4.** Practice Table

9. Follow-on questions

<u>Purpose:</u> Confirm assessor understanding and obtain missing information.

<u>Description:</u> If necessary the assessor asks follow-on questions. After all participants have provided their votes' explanation the assessor might ask follow on questions or seek clarification to some answers. In interest of time, the assessor should keep this short. The completion of all entries in the Process Table serves as exit criteria for the task. If there are items in which the assessor wants to go deeper, the assessor should make a note to retake the conversation at a later time and move on.

10. Definitive vote

<u>Purpose:</u> Avoid voting errors due to misinformation, misunderstanding or unequal information and provide closure a sense of closure.

Description: At this point the assessor will ask participants to take a final vote. This vote has the effect of transforming individual judgments into a collective decision, bringing a sense of closure and accomplishment to the participants. Although vote changes might affect the practice rating, the assessor must make a note in the case of misinformation and unequal information as a process weakness.

11. Vote recording

Purpose: Collect evidence.

Description: Participants record their vote in the Vote Table, see **Fig. 5**. Each participant has its own form to vote and, of course, the forms are not attributable to a particular participant. The purpose of recording the votes is twofold: 1) to have a backup if any of the findings is challenged and 2) to provide an indication of the validity and strength of the findings to those that did not take part in the interview. For example, a finding where 90% of the interviewees voted "seldom done" or "never" it is easier to accept and would trigger different improvements actions, than one where 80% of the participants say it is practiced "most of the time" and the other 20% say they "don't know".

**Fig. 5.** Vote Table. Each row corresponds to a user story/specific practice in the respective process area

12. Validate preliminary findings

Purpose: Confirm the assessor's understanding of the state of the practice and correct factual mistakes.

Description: Instead of waiting until the end of the interview or later, to confirm a batch of preliminary findings like prescribed in the SCAMPI approach, the proposed interview process includes a quick validation step at the end of each iteration to confirm the assessor's understanding of the state of the practice. Because this takes place in the context of what is being discussed and what was said is still vivid in the minds of the interviewees, the possibility of misreading the situation with the consequent frustration and rework is avoided.

First the assessor will make a quick judgment of whether the practice is Fully Implemented (FI), Largely Implemented (LI), Partially Implemented (PI), or Not Implemented (NI), with the help of the rules in **Table 3** and the information collected in the Practice Table. The assessor will then explain his conclusion using the reasoning behind the voting rules and paraphrasing the information gathered through explanations and follow-on questions. Factual misunderstandings are corrected on the spot and matters of interpretation put on a parking lot for further discussion at a later time. The group then moves to the next step.

**Table 3.** Vote interpretation rules

| Scenario | Rating | Reasoning |
|---|---|---|
| All the participants vote "Always" or "Most of the time" ("Strongly Agree" or "Agree") | Fully Implemented (FI) | All the participants know about the practice and they all perform it to some extent under most circumstances |
| All participants vote "Never" or "Seldom" ("Strongly Disagree" or "Disagree") | Not Implemented (NI) | One or more participants could have tried the practice, the "seldom" votes, in the past or through individual efforts but the practice is not being performed |
| A majority of the participants vote "Always" or "Most of the time" ("Strongly Agree" or "Agree"). The dissenting votes are "Don't know" | Largely Implemented (LI) | Most of the participants are performing the practice and those that don't is because they didn't seem to be aware of them. This could be due to lack of training, weaknesses in the onboarding process or lack of an organizational level policy |
| A majority of the votes fell in the "Seldom", "Most of the time" and "Don't know" categories | Partially Implemented (PI) | This clearly points to a practice that is carried out through individual efforts with some success, the "most of the time" votes, but is not institutionalized as indicated by the "seldom" and "Don't know" votes |
| Other | Assessor judgment | |

13. Assessor makes a determination about what to do next

Purpose: Avoid wasting time and embarrassing participants by pursuing dead end lines of inquiry.

Description: The assessor decides if it is worth continue exploring the same process area or move to the next. Normally the assessment will move from one user story to the next within the same process area and once all user stories have been assessed, to the next process area. Sometimes however, after the exploration of a few user stories, it might become obvious the assessed organization does not meet the intent of the process area, and is of no use and almost demeaning to continue asking questions for which the answer is already known. At this point, the assessor should ask the group whether it is worth continuing with the current process area or move to the next.

14. Review and allocate parking lot issues

Purpose: Eliminate items that no longer need to be revisited and assign responsibility for collecting further evidence.

Description: To conclude the group interview, participants review items put in the parking lot. Some of those would have probably resolved themselves through explanations subsequent to the decision that put them there in the first place. Unresolved issues, are assigned to specific participants to gather additional evidence, most likely in the form of work products or descriptions of alternate practices. A meeting with the group is scheduled for the next day.

15. Close remaining parking lot issues

Purpose: To dispose of outstanding issues.

Description: All outstanding parking lot issues are disposed of. Some items might not have a single best answer and in this case to avoid damaging the relation between the assessor and the interviewees the second best alternative is to agree to disagree. If consensus cannot be reached, the assessor in his character of expert has the last word on the disposition of the item but has to leave established that consensus was not reached.

## 5 Final findings

After the group interview process is concluded and all pending issues disposed, the assessor rates the specific goals, determines whether each process area is satisfied or not and derives strengths and weaknesses from the practitioners and management affirmations and his own observations[1]. Optionally an unofficial maturity level might be reported.

Final findings are goal-level statements that summarize the gaps in process area implementation [13]. Strengths are enablers of organizational development. Implementations worth highlighting might be included in the final findings as long as they don't seem to be there just to have something to say on the "bright side". Weaknesses are inefficient implementations of a key practice or hurdles to be overcome to make the improvement initiative successful.

The judgments made about goal satisfaction are driven by the validated preliminary findings and the observations of the assessor. When a goal is not satisfied, it is important to be able to describe how the set of documented weaknesses or the extent of implementation of the associated practices led to this rating. It is also important to link this rating to one or more problems experienced by the organization to make, a compelling case for improvement.

---

[1] In this we differ from SCAMPI which tries to be totally data driven. We believe the experience of the assessor is relevant especially in a process improvement setting.

## 6    Management Interview

Although the focus of this paper is the use of the nominal group interview technique, it would not be complete without a brief description of the management interviews that complemented it.

The management interviews held included meetings with middle and senior managers as well as with user representatives. They touched on the perceived problems, the improvement goals, the organization culture, the political situation and the participants' opinion about the improvement initiative. The main purpose of these meetings was to collect information that would be helpful in the development of a viable improvement plan. The secondary was to give an opportunity to everybody to be heard, a precondition for buying-in into whatever was going to be proposed later.

The interviews were semi-structured and guided by the open-ended question shown by **Fig. 6**. The semi-structured format was considered appropriate for two reasons: 1) because of the limited budget and availability of the participants, the meetings were fundamentally exploratory and as such, the questions were expected to evolve not only between interviews but also while performing them; 2) because of the diversity of stakeholders the questions and the order in which they were asked was modified to reflect each interviewee's responsibility and perspective. In total seventeen of this meetings were conducted.

## 7    Experience

The group interview technique described here was employed twice in the course of assessing the organization which has development sites at two different locations. In both cases the reaction to it was much the same which gives the technique some extra credibility over a single data point case.

At the first location the number of participants was four and at the second seven. Participants received no special training nor were they require to prepare artifacts before the interviews. As shown by the flowchart in **Fig. 1**, any explanation needed was provided "just in time" as part of the assessment process.

Each group interview consisted of two three and a half hours sessions. In one case the sessions were scheduled in two different days, in the other we had a morning and an afternoon session. During the sessions there were little or no signs of fatigue. The use of the cards created a lively environment which was marked by the anticipation of knowing how the others would vote after each user story was presented.

Everybody present at the interview participated, even those that because of personality or opinion, were reluctant in the beginning. In this regard, I just can speculate as to the why. For those normally withdrawn, the engagement was perhaps the result of having the opportunity to talk and be listened to. For others the possibility of change that the assessment opened up. Those that thought the assessment was a bad idea, were put in an uncomfortable position by the round-robin mechanism which left them with no choice but to decline to talk and be perceived as negative and childish or participate.

Participating when you did not believe in it though, would trigger a feeling of dissonance which could, unconsciously, be resolved by saying to oneself that this type of assessment was not so bad, and fostering engagement.

Whatever the reason, engagement was achieved within a couple of voting rounds and maintained through the assessment. These observations are consistent with those mentioned by Delbecq et al [5] and also Gresham [14] in his dissertation "Expressed Satisfaction with the Nominal Group Technique Among Change Agents" and Haugen [15] in his study of the Planning Poker.

---

Current situation

- What is your organization responsibility with regards to software development?
- The 2013 User Committee Report identified a number of problems: communications with user, prioritization, performance and usability, lack of predictability, third party participation. Some of the same problems repeat in the 2014 Report. Do you agree with these problems?
- Do these problems affect your funding, the survival of the organization, why is important to solve them?
- Are there other pain points not mentioned in the reports?
- What do you think is the root cause of these problems?
- What do you see as impediments to solve this problems?
- Do other members of the management team share your assessments?

Environment

- What are your improvement goals? How would you know you have reached them?
- If you were to establish development processes or ask team members to report time or status, how do you think they would react to that?
- Is there any organizational policy mandating software development, project management, quality assurance? Why not?
- Does management provide adequate funding, physical facilities, skilled people, training and appropriate tools to perform the processes?
- Do you assign responsibility and authority for performing the process, for example through job descriptions?

Closing

- Before we close the interview, is there anything you would like to add, any points we missed and you would like to comment on?

**Fig. 6.** Management interview guide

## 8   Summary

In this paper we described the successful use of a novel process assessment method based on the Nominal Group Technique and the Planning Poker ceremonies.

The method application was successful in the sense that not only correctly identified areas of improvement but also played a reconciliation role among groups with different views to the point, that people that was originally against the assessment ended up being very supportive. Furthermore, it did so in an unobtrusive and economical way.

It is worth mentioning here, that the premise on which the method is based, is that all interviewees will answer truthfully and to the best of their knowledge. This is a reasonable expectation in the case of an assessment with the purpose of improvement, but not so in the case of a process evaluations for source selection or contract qualification, reason for which at this time the author would not recommend its application in those context without further research.